# Enabling Reusability in Agile Software Development

Sukhpal Singh
M.E. (S.E.)
Thapar University, Patiala

Inderveer Chana
Ph.D. (C.S.E.)
Thapar University, Patiala

## ABSTRACT
Software Engineering Discipline is constantly achieving momentum from past two decades. In last decade, remarkable progress has been observed. New process models that are introduced from time to time in order to keep pace with multidimensional demands of the industry. New software development paradigms are finding its place in industry such as Agile Software Development, Reuse based Development and Component based Development. But different software development models fail to satisfy many needs of software industry. As aim of all the process models is same, i.e., to get quality product, reduce time of development, productivity improvement and reduction in cost. Still, no single process model is complete in itself. Software industry is moving towards Agile Software Development. Agile development does not obviously fit well for building reusable artifacts. However, with careful attention, and important modifications made to agile processes, it may be possible to successfully adapt and put on agile methods to development of reusable objects. The model being proposed here combines the features of Agile Software Development and reusability.

## 1. INTRODUCTION
Essence of Agile Software Development is rapid development and less cost. Thus, it somewhere compromises with quality and also unable to provide reusability of its developed components. Agile Software Development provides specific solutions whereas Reuse and Component based Development have faith in generalized solutions. We need reusable artifacts to attain reusability. Reusable artifacts are code and other components (analysis and design documents, patterns, etc.) that can be reused from one project to another. In order to create components that are reusable, a big-depiction view must be taken while they are being developed, rather than simply aiming on the current use. What other types of systems might be able to benefit from this component? How many different ways might one want to use it? What are the requirements of the domain, as compare to simply this application in the domain? These are a few of the questions that must be asked when thinking about making components reusable and more general-purpose. When developing reusable artifacts, agile development's Documentation, Software Quality Assurance, Application-Specific Development, and Continuous Redesign assumptions may not be valid.

### 1.1 Agile Software Development
Agile Software Development (ASD) methods and techniques are being followed in the industry from the last decade to get quality product and to reduce development time. Rapid development and ability to accommodate changes at any level of development gives the competitive advantage to the agile processes over traditional processes. Bring comparatively new to software engineering; research on agile processes is going on as to combine light-weight processes and other processes. Recent research shows that only limitation of ASD is its inability to reuse components those are developed through agile processes. On the whole rapid software development ignores reusability. Japanese projects also exhibited higher levels of reuse while spending more time on product design as compared to American teams which spend more time on actual coding and concludes that Indian firms are doing great job in combining conventional best practices, such as specification and review, with more flexible techniques that should enable them to respond more effectively to customer demands. If such a trend is replicated across the broader population, it suggests the Indian software industry is likely to experience continued growth and success in future [1].

### 1.2 Evolution of Software Development Processes
The evolution of software development processes have been summarized in Fig. 1. Waterfall model was being followed where requirements are fixed and the next phase starts when the earlier one finished. It's the representative of the traditional methods [13]. To overcome the limitations of waterfall model evolutionary model and spiral model comes into picture where prototype is first made and then that is converted to the working software. But all have one common limitation that no process could handle the change of requirements at later phases. Agile development which includes many methodologies as XP, SCRUM, Lean Software Development, Feature Driven Development (FDM), Test Driven Development etc. is being accepted in industry because of adaptation to change even at the later stages of the development and also for rapid development [2].

### 1.3 Reusability
In computer science and software engineering, reusability is the likelihood a segment of source code that can be used again to add new functionalities with slight or no modification. Reusability modules and classes reduce implementation time, increases the probability that prior testing and use has eliminated bugs and localizes code modifications when a change in implementation is required [19].

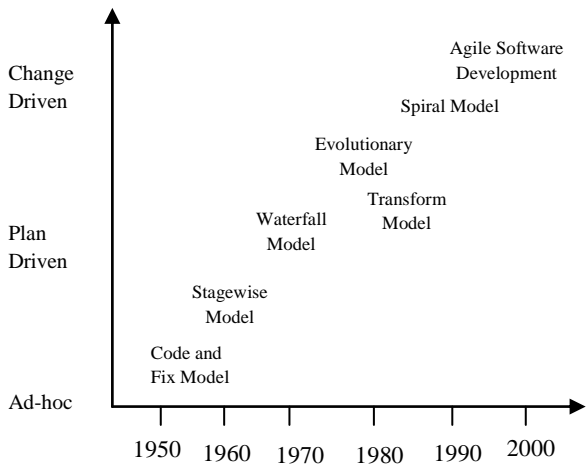

**Fig. 1: The Evolution of Software Process Models**

## 2. LITERATURE REVIEW
### 2.1 Philosophies for Agile Software Development

A common ground for agile software development was defined in 2001, when 17 experienced and recognized software development "gurus", inventors and practitioners of different agile software development methods gathered together. Participants agreed and signed The Manifesto for Agile Software Development [3]. This manifesto declares the main values of agile software development [6]: "We are uncovering better ways of developing software by doing it and helping others does it. Through this work we have come to value:

• Individuals and interactions over processes and tools.
• Working software over comprehensive documentation.
• Customer collaboration over contract negotiation.
• Responding to change over following a plan.

That is, while there is value in the items on the right, we value the items on the left more" [4].

### 2.2 Agile Methodologies

There are a variety of software development processes that currently claim to be agile. Space does not allow us to give an overview of all of the agile processes we have reviewed. However, since Extreme Programming (XP) is probably the most well-known agile process (Beck, 2000; Strigel, 2001), we use it to illustrate representative agile process concepts [10].

*2.2.1 Extreme Programming (XP)*

It can be argued that the popularity of XP helped pave the way for other agile processes. Kent Beck, one of the chief architects of XP [7], states that XP is a "lightweight" development method that is tolerant of changes in requirements [17]. It is "extreme" in that "XP takes commonsense principles and practices to extreme levels". XP is based on the following values:

Communication and Feedback: Face-to-face and frequent communication among developers and between developers and customers is important to the "health" of the project and the products under development. Feedback, through delivery of working code increments at frequent intervals, is also considered critical to the production of software that satisfies customer needs.

Simplicity: XP assumes that it is more efficient to develop software for current needs rather than attempt to design flexible and reusable solutions. Under such an assumption, developers pursue the simplest solutions that satisfy current needs.

Responsibility: The responsibility of producing high-quality code rests ultimately with the developers.

XP consists of technical and managerial practices that are integrated in a complementary manner. The architects of XP take great care to point out that the individual techniques and practices of XP are not new; it is the manner in which they are woven together that is unique. They also stress that the techniques and practices have proven their worth in industrial software development environments [9].

*2.2.2 XP Process and Practices*

The four core activities [17] given by Roger S. Pressman have been summarized in Table 2.1.

**Table 2.1: The four core activities of XP**

| Activities | Purpose |
| --- | --- |
| Planning | It begins with the creation of stories that describe required features and functionality for software to be built |
| Design | It provides implementation guidance for a story as it is written (Refactoring). |
| Coding | XP recommends that two people work together to at one computer workstation to create a code for a story (Pair programming). |
| Testing | Test cases are specified by customer and focus on overall system features and functionality that are visible and reviewable by customer. |

Pair programming [11], one of the more well-known XP practices, is a technique in which two programmers work together to develop a single piece of code (Fraser et al, 2000; Williams et al, 2000; Williams & Upchurch, 2001).

Refactoring occurs when a change to the internal structure of a system preserves the externally observable functionality of the system [17]. Refactoring is especially effective when large changes can be decomposed into smaller steps that can be carried out using refactoring's that have been developed by fowler and others.

*2.2.3 Agile Approaches*

Refactoring: Refactoring is a fundamental to Agile Development. Refactoring is a development process for restructuring an existing code, altering its internal structure without changing its external behaviour [17]. It's a process of improvement to an existing software artefact. It improves the design of the software and makes software easier to understand. It helps to avoid errors and to maintain, and modify a program with more accuracy and speed. Code duplication is the main cause for bad smells in code. Eclipses (java), Jbuilder (Java), ReSharper (.NET), and Refactor for Visual Basic are a few names [8].

TDD (Test Driven Development): TDD is one of the most profound agile development practices. It TDD is an evolutionary approach to development which instructs to have TFD (Test First Design) intent. TDD actually helps to meet deadlines by eliminating debugging time, minimizing design speculation and re-work, and reducing the cost and fear of changing working code [12]. TDD starts with writing a test to fail and then coding is being done to pass that test. If written code pass the test then code refactoring is being done otherwise again the code is being written and tested. It's a way to do unit testing. The cycle is repeated till the dead end. Unit framework family of open source tools is a very common used tool support for TDD in agile development. TDD is basically the combination of Test First Design and Refactoring [5]. The steps of Test first Design has been summarized in Fig. 2 in the form of flowchart.

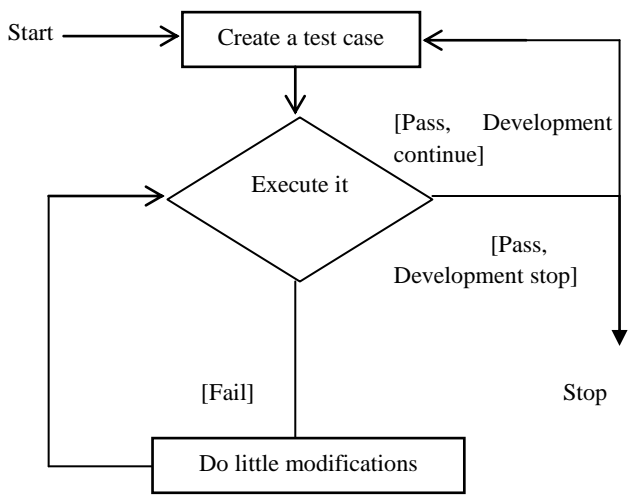

**Fig. 2: Steps of Test First Design**

## 2.3 Agile Software Development and SPI

One key research area related to agile processes is in software process improvement (SPI). The differences [14] between software improvement initiatives of Traditional and Agile development have been summarized in Table 2.2.

**Table 2.2: Underlying differences of Traditional and Agile Software Development and SPI**

| Criteria | Traditional Software Development and SPI | Agile Software Development and SPI |
|---|---|---|
| Process control | Control on organization level | Self-organizing teams |
| Primary means of knowledge transfer | Document based knowledge transfer | Face to face communication |
| Immediate focus of process improvement | Improvement of organisational software development process | Improvement of daily working practices of ongoing projects |
| Software development Process | Universal approach and repeatable solution to provide high assurance | Flexible approach adapted with collective understanding of contextual needs to provide faster development times, responsiveness to rapid changes, Increased customer satisfaction and lower defect rates. |

## 2.4 Reusability

To share code between different applications is considered to be the reusability but a variety of assets [15] can be reused across software development processes are summarized in Table 2.3.

**Table 2.3: Possible Assets for Reuse**

| Intermediate Artefacts | Implemented Artefacts | Project Management and Quality Assurance Artefacts |
|---|---|---|
| Requirements | System | Process models |
| Architectures | Frameworks, components, modules, packages | Planning models |
| Designs | UML models, interfaces, patterns | Cost models |
| Algorithms | Libraries | Review and inspection forms |
| Documentations | Test cases | Analysis models |

Reusability increases not only the productivity of the developers but also the reliability and maintainability of the software products. Many software companies have repository to support the reusability. Object-Orientation also offers reusability. No of techniques are available to support reusability. Considerable research and development is going on in reuse; industry standards like CORBA have been created for component interaction; and much domain specific architecture, toolkits, application generators and other related products that support reuse and open systems have been developed [16].

## 2.5 Reusability in Agile Software Development

There are three ways or technologies discussed one by one below, by which reusability can be incorporated in agile software development.

### 2.5.1 Component Based Development (CBD)

CBD is a reusability approach that can be found in Microsoft .NET Framework and J2EE (Java2 Enterprise Edition). Component Based Software Engineering (CBSE) process identifies not only candidate components but also qualifies each components interface, adapts components to remove architectural mismatches, assembles components into selected architectural style, and updates components [20].

### 2.5.2 Refactoring to Design Patterns

To provide a software system quality in terms of reusability, flexibility and extendibility, refactoring is a significant solution [5]. Use of design patterns in an application increases reusability and maintainability. One new emerging approach that refactoring to design patterns seems to have promising future in reusability discipline. Research is going on in the field of pattern mining as to find new approaches and new tools. Refactoring has been gained much more attention in the object oriented software development. Refactoring to Patterns suggests that using patterns to improve an existing design is better than using patterns early in a new design. We improve designs with patterns by applying sequences of low-level design transformations. A design pattern is not a finished design that can be transformed directly into code. It is a description or template for how to solve a problem that can be used in many different situations. Object oriented design patterns typically show relationship, interactions between classes or objects, without specifying the final application classes or objects that are involved.

### 2.5.3 Reusable Architectures

Reusable architectures can be developed from reusable architectural patterns [18] as in FIM architecture [19] which operates at three different levels of reuse: Federation, domain and application. They focus on how non-functional property reusability relates to the software architecture of a system. They presented a suggested software process model for reuse based software development approach.

## 2.6 Risk Analysis

From the past we learnt that any new development has some associated risks. Major risk is where the decisions are made on the vision and on the future scope of the project. Although its probability is low because we are considering that the team members are experts but its impact is high on project and on the organization itself. Another risk is technical risk which is a common to all of the organizations that what if the technology changes. A new technology comes in the market. Risk related to design patterns is of time as sometimes more than one design pattern seems to be the solution. Risk related to repository is that the assets are kept there which are not being used since a long time. It's related to the maintenance of the repository. The impact is low if the repository is small but as the repository grows the impact will also grow as the time to find the asset will increase.

## 2.7 Limited Support for Building Reusable Artifacts in Agile

Reusable artifacts are code and other components (analysis and design documents, patterns, etc.) that can be reused from one project to another, in their entirety or at least in a major part. In order to create components that are reusable, a big-picture view must be taken while they are being developed, rather than simply focusing on the current application [21]. This separation of the product-specific development environment from the reusable artifact development environment is a primary feature of the reuse-oriented framework called the Experience Factory developed by researchers at the University of Maryland at College Park (Basili, Caldiera, & Rombach, 1994). Continuous redesign is difficult when we are not developing application-specific artifacts.

## 3. PROPOSED SOLUTION

A model capable of classifying, storing, searching and retrieving the components from the agile repository by using pattern matching algorithms has been proposed in this work is shown in Fig. 3. There are some different storage and retrieval methods are available for the classification of components in software library. This model will help to make searching faster based on classification of components. The flowchart of proposed model is shown in Fig. 4.

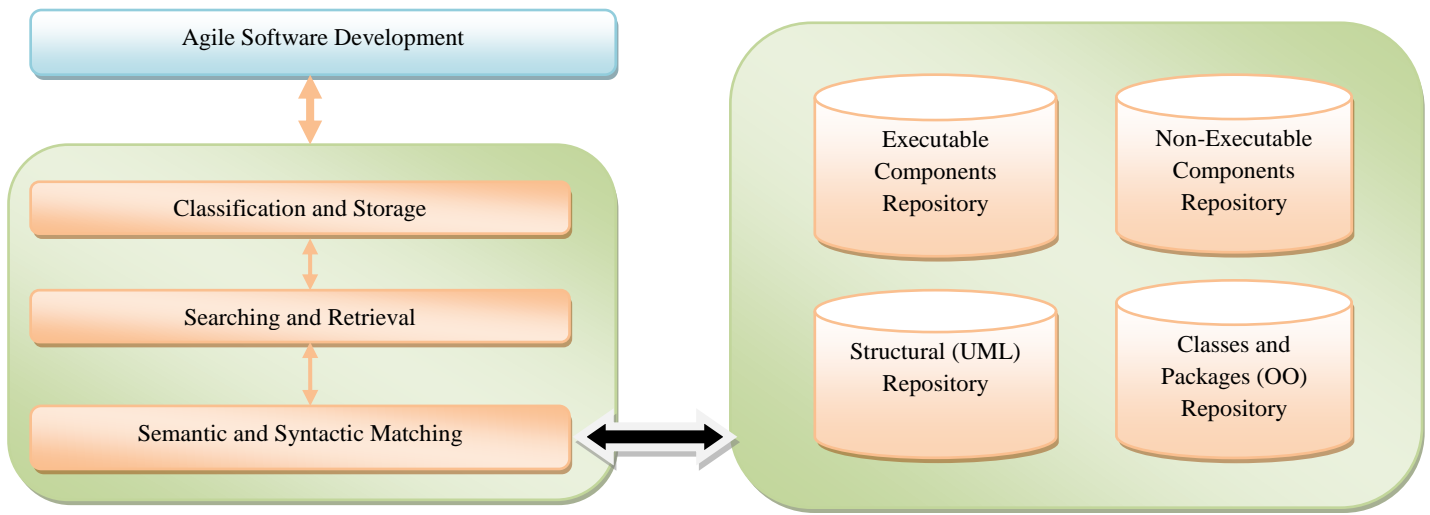

**Fig. 3: Proposed Model**

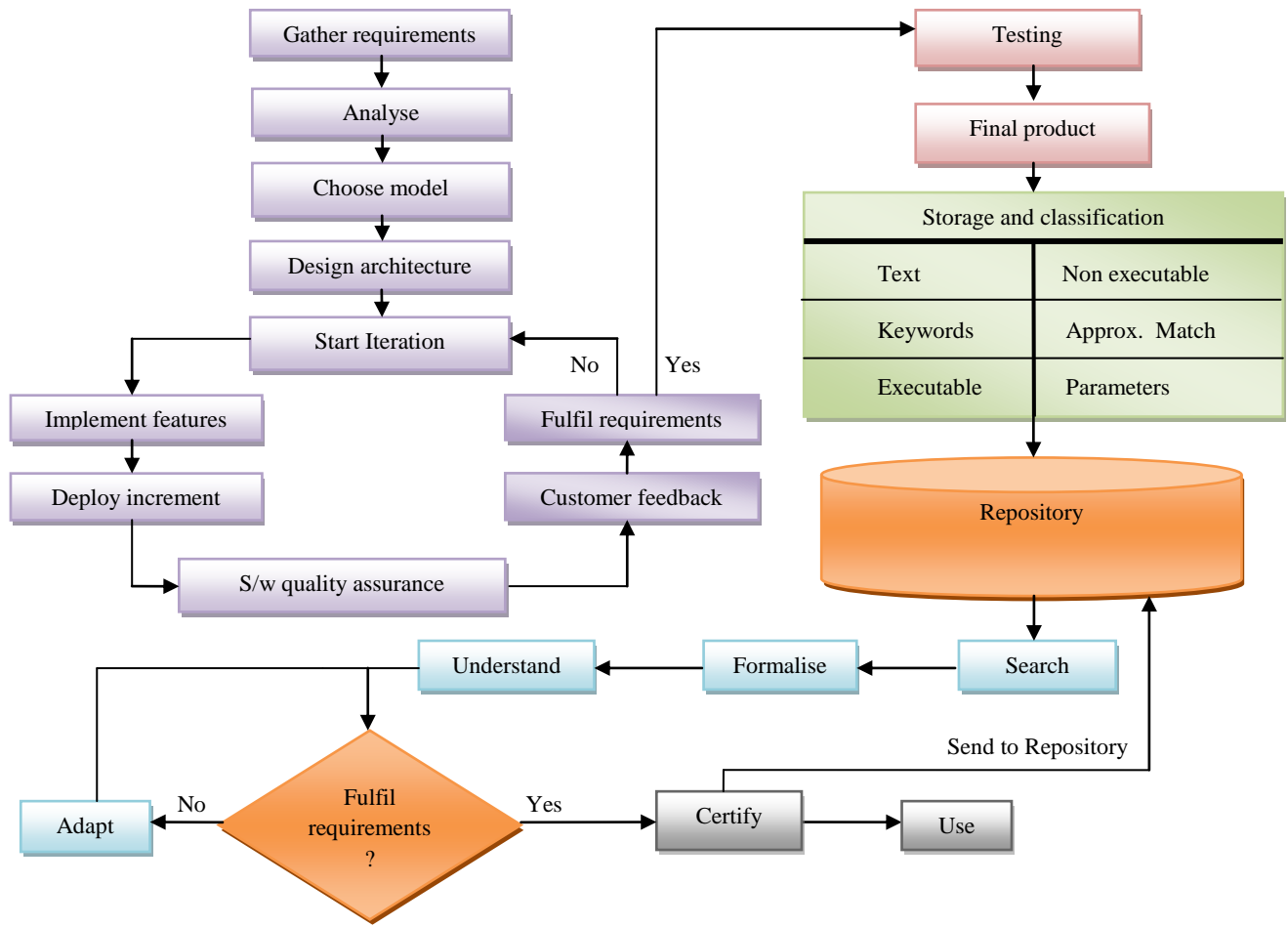

**Fig. 4: Flowchart of Proposed Model**

## 3.1 Storage and retrieval criteria

### 3.1.1 Information retrieval methods
These are methods that depend on a textual analysis of software assets. It is important to acknowledge that the storage and retrieval of software assets is nothing but a specialized instance of information storage and retrieval. The registration component is retrieved with other information by using text "software reuse" is shown in Table 3.1.

**Table 3.1: Informational retrieval methods**

| Component name | Component id | Label | Language |
|---|---|---|---|
| registration | Text_6562 | software reuse | C++ |

### 3.1.2 Descriptive methods
These are methods that depend on a textual description of software asset. While information retrieval methods represent assets by some form of text, descriptive methods rely on an abstract surrogate of the asset, typically a set of keywords, or a set of facet definitions. The feedback component is retrieved with other information by using keyword "Agility" is shown in Table 3.2.

**Table 3.2: Descriptive methods**

| Component name | Component id | Keyword | Language |
|---|---|---|---|
| feedback | Key_6522 | Agility | Java |

### 3.1.3 Operational semantics methods
These are methods that depend on the operational semantics of software assets. This can be applied to executable code, and proceed by matching candidate assets against a user query on the basis of the candidates behaviour on sample inputs. The update.exe executable component is retrieved with other information by writing the executable name "update" is shown in Table 3.3.

**Table 3.3: Operational semantics methods**

| Component name | Component id | Executable |
|---|---|---|
| update.exe | Exe_4329 | update |

*3.1.4 Denotational semantics methods*
These are methods that depend on the denotational semantic definition of software assets are shown in Table 3.4. Unlike operational methods, they can also be applied to non-executable assets (such as specifications). The initial requirements non-executable component is retrieved with other information by writing the non-executable name "requirements" is shown in Table 3.4.

**Table 3.4: Denotational semantics methods**

| Component name | Component id | Non Executable |
|---|---|---|
| Initial requirements | nonExe_7215 | requirements |

*3.1.5 Topological methods*
The main feature of topological methods is their goal, which is to identify library assets that minimize some measure of distance to the user query. The Support component is retrieved with other information by writing the identity "port" is shown in Table 3.5.

**Table 3.5: Topological methods**

| Component name | Component id | Identity |
|---|---|---|
| Support | Id_1213 | port |

*3.1.6 Structural methods*
The main discriminating feature of structural methods is the nature of the software asset they are dealing with: typically, they do not retrieve executable code, but rather program patterns, which are subsequently instantiated to fit the user's needs. The class name city is retrieved with other information by writing the package name "state" is shown in Table 3.6.

**Table 3.6: Structural methods**

| Package | Class | Pattern | Language |
|---|---|---|---|
| State | City | Object oriented | Java |

*3.1.7 Analysis of accessing criteria*
The technical, managerial and human criteria have been summarized in Table 3.7 and Table 3.8. The abbreviations used in these tables are VL=Very Low, L=Low, M= Medium, H= High, VH= Very High. The Analysis of retrieving criteria has been shown in Fig. 5.

**Table 3.7: Summary of Technical criteria**

| | Technical | | | | | |
|---|---|---|---|---|---|---|
| Methods | Precision | Recall | Coverage ratio | Time complexity | Logical complexity | Automation |
| Informational | M | H | L | L | M | H |
| Descriptive | H | H | VH | VL | L | VH |
| Operational | VH | H | H | M | M | VH |
| Denotational | VH | H | H | VH | VH | M |
| Topological | U | U | VH | H | M | H |
| Structural | VH | VH | VH | VL | L | VH |

**Table 3.8: Summary of managerial and human criteria**

| | Managerial | | | | Human | |
|---|---|---|---|---|---|---|
| Methods | Investment cost | Operational cost | Pervasiveness | State of development | Difficulty of use | Transparency |
| Informational | VL | L | H | H | M | H |
| Descriptive | H | H | H | H | VL | VH |
| Operational | L | M | M | M | L | VH |
| Denotational | H | H | L | L | M | M |
| Topological | VH | VH | L | L | VH | VH |
| Structural | M | L | L | L | VL | VL |

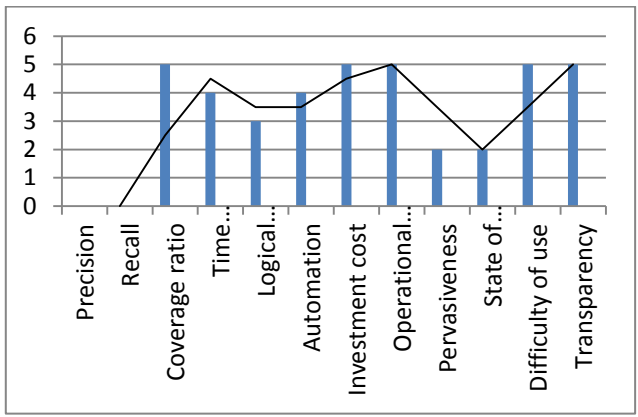

**Fig. 5: Analysis of accessing criteria**

## 4. ADVANTAGE OF PROPOSED APPROACH

Essence of Agile Software Development is rapid software development with reduced overheads. This proposed model will help to make searching faster based on classification of components and introducing reusability in Agile Software Development. It will be accepted widely if pattern based architecture designing, design patterns, UML based analysis and designing is incorporated. Pattern based architecture oriented agile development and use of OO patterns as refactoring to design patterns will make a space for reusability and reusable artifacts.

### 4.1 Related Work

We had conducted a survey on the number of approaches existing for Agile Software Development [7, 17] and Reusability [19] individually, but the proposed model combines both Agile Software Development and Reusability into a single approach for achieving efficient classification, storage and retrieval of software components. Presently there is no such approach as presented in proposed model which combines Agile Software Development and Reusability.

## 5. CONCLUSION AND FUTURE SCOPE

### 5.1 Conclusion

Agile Software Development has encouraging future in the software industry and is capable of fulfilling the requirements of the industry. Thus, at times it compromises with quality and is incapable of providing reusability of its developed modules. Agile Software Development offers particular solutions whereas Reuse and Component based Development believe in generalized solutions. Reuse based software engineering and agile development is an open research area in fast growth.

### 5.2 Future Scope

The future scope of this work is to analyze and to incorporate risk factors in Agile Software Development systematically and find the critical success factors of the agile software development process and also identify the various risk factors using risk analysis of introducing reusability in agile software development and offer a model that will help us to achieve reusability in Agile Software Development. Reusability can also be automated in agile software development using an automated tool.